\RequirePackage{fix-cm}
\documentclass[smallextended]{svjour3}       % onecolumn (second format)
\smartqed  % flush right qed marks, e.g. at end of proof
\usepackage{graphicx}
\usepackage{latexsym,amssymb,amsmath,color,bm,url,natbib}

\newcommand{\R}{\mathbb{R}}

\begin{document}

\title{Investigating consumers' store-choice behavior via hierarchical variable selection
%Analyzing store choice factors based on hierarchical variable selection 
%Analysis of store choice factors using hierarchical variable selection based on mixed-integer optimization
%\thanks{Grants or other notes
%about the article that should go on the front page should be
%placed here. General acknowledgments should be placed at the end of the article.}
}
%\subtitle{Do you have a subtitle?\\ If so, write it here}

%\titlerunning{Short form of title}        % if too long for running head

\author{Toshiki Sato \and Yuichi Takano \and Takanobu Nakahara 
%etc.
}

%\authorrunning{Short form of author list} % if too long for running head

\institute{
Toshiki Sato \at
Graduate School of Systems and Information Engineering, University of Tsukuba, 1-1-1 Tennodai, Tsukuba-shi, Ibaraki 305-8573, Japan
%\emph{Present address:} of F. Author  %  if needed
\and
Yuichi Takano \at
School of Network and Information, Senshu University, 2-1-1 Higashimita, Tama-ku, Kawasaki-shi, Kanagawa 214-8580, Japan %  \\
% \emph{Present address:} of F. Author  %  if needed
\and
Takanobu Nakahara \at
School of Commerce, Senshu University, 2-1-1 Higashimita, Tama-ku, Kawasaki-shi, Kanagawa 214-8580, Japan
% \emph{Present address:} of F. Author  %  if needed
}

\date{Received: date / Accepted: date}
% The correct dates will be entered by the editor

\maketitle

\begin{abstract} % 150 to 250 words in the case of ADAC
This paper is concerned with a store-choice model for investigating consumers' store-choice behavior based on scanner panel data. 
Our store-choice model enables us to evaluate the effects of the consumer/product attributes not only on the consumer's store choice but also on his/her purchase quantity. 
Moreover, we adopt a mixed-integer optimization (MIO) approach to selecting the best set of explanatory variables with which to construct a store-choice model. 
We devise two MIO models for hierarchical variable selection in which the hierarchical structure of product categories is used to enhance the reliability and computational efficiency of the variable selection. 
We assess the effectiveness of our MIO models through computational experiments on actual scanner panel data. 
These experiments are focused on the consumer's choice among three types of stores in Japan: convenience stores, drugstores, and grocery supermarkets. 
The computational results demonstrate that our method has several advantages over the common methods for variable selection, namely, the stepwise method and $L_1$-regularized regression. 
Furthermore, our analysis reveals that convenience stores tend to be chosen because of accessibility, drugstores are chosen for the purchase of specific products at low prices, and grocery supermarkets are chosen for health food products by women with families.

% \PACS{PACS code1 \and PACS code2 \and more}
% \subclass{MSC code1 \and MSC code2 \and more}
\end{abstract}

\section{Introduction}\label{sec:1}
Variable selection, also known as feature/attribute/subset selection, involves selecting a set of relevant explanatory variables from many candidates and using them to construct a statistical model. 
This procedure facilitates interpretation of the subsequent analysis of the statistical model, and enhances the model's predictive performance by preventing overfitting~(\citealt{GuEl03}). 
Because datasets are becoming ever larger, computational methods for variable selection are under active investigation in the fields of machine learning and data mining~(\citealt{BlLa97,GuEl03,KoJo97,LiMo07}). 

A direct way of selecting the best set of explanatory variables is to evaluate all possible subset models~(\citealt{FuWi00}). 
However, this approach is unsuitable in practice unless there are sufficiently few candidate variables. 
Although the stepwise method~(\citealt{Ef60}), regularized/penalized regression~(\citealt{Ti96}), and metaheuristics~(\citealt{Yu09}) are practical approaches to variable selection, they do not necessarily find the best set of explanatory variables under the given goodness-of-fit measures. 
Hence, alternative approaches based on mixed-integer optimization (MIO) are now receiving considerable attention~(\citealt{BeKi17,BeKi16b,KoYa09,MaPe14,MiTa15a,MiTa15b,SaTa16a,SaTa17,UsRu16,WiSa17}) because these have the potential to provide the best set of explanatory variables with respect to several goodness-of-fit measures. 

The purpose of this paper is to use MIO-based variable selection to analyze the factors that affect which stores consumers choose. 
Although there have been various studies of consumer store choice~(\citealt{BaPa02,BlDe98,BrCh09,LeTi02,PaZi06,ReTe09}), only \cite{SaTa16b} used MIO-based variable selection to analyze it. 
Specifically, they used panel data of the barcode scans of a consumer's previous purchases to predict which stores s/he would visit. 
Their predictive model accounted for the different purchasing patterns among the targeted stores, and hence revealed the product set associated with each store. However, \cite{SaTa16b} did not predict how many purchases the consumer would make. 
In addition, they did not take into account the hierarchical structure (or inclusion relation) of product categories when constructing their predictive model. 

Structured regularization is a method whereby existing structural information (e.g., relationships between explanatory variables) is used in the construction of a statistical model. 
For instance, sets of interrelated variables are selected simultaneously in the grouped variable selection, whereas the selection of hierarchical variables is prioritized according to previous information in the hierarchical variable selection~(\citealt{BiTa13,HuZh11,JaOb09,JeAu11,KiXi10,TiSa05,YuLi06,ZhRo09}). 
These studies used either the stepwise method or regularized/penalized regression as the algorithm for selecting the structured variables. 
However, to the best of our knowledge, no existing study has used the MIO-based method to select hierarchical variables. 

Hence, we propose a purchase-quantity-based store-choice model for analyzing the factors involved in consumer store choices. This model allows us to explore a consumer's store choices and his/her purchase quantities simultaneously. To construct the store-choice model, we devise two MIO models for selecting hierarchical variables. We evaluate the effectiveness of each MIO model by comparing its computational performance with those of the stepwise method and $L_1$-regularized regression. We also use computational results from panel data of actual barcode scans to clarify the factors involved in consumer store choice. 

The remainder of the paper is organized as follows. In Sect.~\ref{sec:2}, we present our purchase-quantity-based store-choice model. In Sect.~\ref{sec:3}, we formulate MIO models of variable selection for the store-choice model. In Sect.~\ref{sec:4}, we report the computational results of variable selection, and in Sect.~\ref{sec:5} we conclude with a brief summary. 

\section{Store-choice model}\label{sec:2}
Let us consider two stores: store A (positive example) and store B (negative example). Each sample datum $i=1,2,\ldots,n$ corresponds to a visit of a consumer to one or other of these stores. On the basis of these data, \cite{SaTa16b} developed a binary classification model for predicting which store a consumer would visit. In contrast, to include consumer purchase quantity in the store-choice model, we define the explained variable $y_i$ as follows: 
\begin{align}
y_i := \mbox{(purchase quantity at store A)} - \mbox{(purchase quantity at store B)} \notag
\end{align}
for each sample $i=1,2,\ldots,n$. Note that each sample (or visit) is associated with one or other of the two stores. Therefore, for each sample $i$, $y_i$ is a positive integer if store A is chosen or a negative integer if store B is chosen. 

We consider consumer demographics and categories of purchased products as explanatory variables that influence store choice. A set of consumer demographics is denoted by $G_0$, and the value $x_{ij}$ for $j \in G_0$ is given by the corresponding consumer demographic in each sample $i=1,2,\ldots,n$. Sets of large, medium, and small product categories are denoted by $G_1$, $G_2$, and $G_3$, respectively, which are all dummy variables. For all $j \in G_1 \cup G_2 \cup G_3$, $x_{ij} := 1$ if products in category $j$ are purchased in sample $i$; otherwise, $x_{ij} := 0$. The set of all candidate explanatory variables is denoted by $G := G_0 \cup G_1 \cup G_2 \cup G_3$. 

On the basis of these explanatory and explained variables, we consider the following linear regression model: 
\begin{align} \notag
y_i = b + \sum_{j \in G} a_j x_{ij} + \varepsilon_i \quad (i = 1,2,\ldots,n),
\end{align}
where $b$ is an intercept term to be estimated, $a_j$ is a regression coefficient to be estimated for the $j$th explanatory variable, and $\varepsilon_i$ is a prediction residual for each sample $i = 1,2,\ldots,n$. We refer to this regression model as a store-choice model; it explains the effects of the consumer/product attributes not only on the consumer's store choice but also on his/her purchase quantity. 

\section{Mixed-integer optimization models for variable selection}\label{sec:3}
In this section, we firstly present a basic MIO model for variable selection~(\citealt{BeKi16b,KoYa09}). We then describe our MIO models for hierarchical variable selection in the store-choice model. 

\subsection{Basic MIO model for variable selection}
We begin by explaining the decision variables with which we formulate the MIO models. Specifically, $\bm{a} := (a_j)_{j \in G} \in \R^p$ denotes a vector of decision variables representing regression coefficients, where $p$ is the number of candidate explanatory variables, that is, $p := |G|$. Next, $\bm{z} := (z_j)_{j \in G} \in \{0,1\}^p$ denotes a vector of 0--1 decision variables for variable selection; that is, $z_j = 1$ if the $j$th explanatory variable is selected; otherwise, $z_j = 0$. If $z_j = 0$, then we eliminate the $j$th explanatory variable from the regression model by setting its coefficient, $a_j$, to zero. 

We minimize the residual sum of squares subject to the constraint of the upper bound on the number of selected explanatory variables. Consequently, the basic MIO model for variable selection is posed as follows~(\citealt{BeKi16b,KoYa09}): 
\begin{align}
\mathop{\mbox{minimize}}_{\bm{a},\,b,\,\bm{z}} 
& \quad \sum_{i=1}^n \left(y_i - \left( b + \sum_{j \in G} a_j x_{ij} \right) \right)^{\!\!\! 2}  \label{obj:MIQO1} \\
\mbox{subject~to} 
& \quad z_j = 0~\Rightarrow~a_j = 0 \quad (j \in G), \label{con1:MIQO1}\\
& \quad \sum_{j \in G} z_j \le s, \label{con2:MIQO1} \\
& \quad z_j \in \{0,1\} \quad (j \in G), \label{con3:MIQO1}
\end{align}
where $s$ is a user-defined parameter representing the upper bound on the number of selected explanatory variables. The logical implication in~\eqref{con1:MIQO1} can be formulated by using a constraint in the form of a special ordered set of type 1 (SOS1), which is supported by standard MIO software. This constraint implies that no more than one element in the set can have a non-zero value. Therefore, the logical implication in~\eqref{con1:MIQO1} is equivalent to imposing the SOS1 constraints on $\{1 - z_j, a_j\}$ for all $j \in G$. Indeed, if $z_j = 0$, then $1 - z_j$ is non-zero and $a_j$ must be zero from the SOS1 constraints. 

\subsection{MIO models for hierarchical variable selection}
To enhance the reliability of the regression analysis, we exploit the hierarchical (or inclusion) relationship among product categories. For instance, the small category ``seasoning'' is contained in the medium category ``processed foods,'' and these two categories are contained in the large category ``food.'' 

Let $H$ be the set of 3-tuples $(j_1,j_2,j_3) \in G_1 \times G_2 \times G_3$ of product categories having such a hierarchical relationship. In other words, $(j_1,j_2,j_3) \in H$ means that large category $j_1$ contains medium category $j_2$, and medium category $j_2$ contains small category $j_3$. On the basis of these hierarchical relationships, we consider the following constraints in the variable selection: 
\begin{align}
\mbox{Strong hierarchical constraints:} & \quad z_{j_1} \ge z_{j_2} \ge z_{j_3} \quad ((j_1,j_2,j_3) \in H), \notag \\
\mbox{Weak hierarchical constraints:} & \quad z_{j_1} \ge z_{j_2},~z_{j_1} + z_{j_2} \ge z_{j_3} \quad ((j_1,j_2,j_3) \in H). \notag
\end{align}

To gain a better understanding of these constraints, we suppose that for $(j_1,j_2,j_3) \in H$, small category $j_3$ is selected as an explanatory variable (i.e., $z_{j_3} = 1$). In that case, we must select both large category $j_1$ and medium category $j_2$ (i.e., $z_{j_1} = z_{j_2} = 1$) when the strong hierarchical constraints are imposed. In contrast, the weak hierarchical constraints require us to select at least one of the categories that is superordinate to the selected one. Indeed, the weak hierarchical constraints are satisfied by selecting large category $j_1$ (i.e., $z_{j_1} = 1$) even if medium category $j_2$ is not selected. 

These hierarchical constraints can be validated based on the following observations: 
\begin{description}
\item[\bf Reliability of variable selection.]
When a certain product category is a store-choice factor, its superordinate categories are likely to affect the store choice. For this reason, the hierarchical constraints can improve the reliability of selected explanatory variables. 
\item[\bf Accuracy of coefficient estimates.]
Superordinate categories are preferentially selected by the hierarchical constraints. Since such categories involve many samples, the accuracy of the coefficient estimates can be raised. 
\item[\bf Efficiency of MIO computations. ]
The number of feasible subsets of explanatory variables is greatly reduced by the hierarchical constraints. As a result, the MIO computations can be made more efficient. 
\end{description}

The strong hierarchical MIO model is formulated by appending the strong hierarchical constraints to the basic MIO model~\eqref{obj:MIQO1}--\eqref{con3:MIQO1} as follows:
\begin{align}
\mathop{\mbox{minimize}}_{\bm{a},\,b,\,\bm{z}} 
& \quad \sum_{i=1}^n \left(y_i - \left( b + \sum_{j \in G} a_j x_{ij} \right) \right)^{\!\!\! 2}  \label{obj:MIQO2} \\
\mbox{subject~to} 
& \quad z_j = 0~\Rightarrow~a_j = 0 \quad (j \in G), \label{con1:MIQO2}\\
& \quad \sum_{j \in G} z_j \le s, \label{con2:MIQO2} \\
& \quad z_{j_1} \ge z_{j_2} \ge z_{j_3} \quad ((j_1,j_2,j_3) \in H), \label{con3:MIQO2}\\
& \quad z_j \in \{0,1\} \quad (j \in G). \label{con4:MIQO2}
\end{align}
Similarly, the weak hierarchical MIO model is framed as follows: 
\begin{align}
\mathop{\mbox{minimize}}_{\bm{a},\,b,\,\bm{z}} 
& \quad \sum_{i=1}^n \left(y_i - \left( b + \sum_{j \in G} a_j x_{ij} \right) \right)^{\!\!\! 2}  \label{obj:MIQO3} \\
\mbox{subject~to} 
& \quad z_j = 0~\Rightarrow~a_j = 0 \quad (j \in G), \label{con1:MIQO3}\\
& \quad \sum_{j \in G} z_j \le s, \label{con2:MIQO3} \\
& \quad z_{j_1} \ge z_{j_2},~z_{j_1} + z_{j_2} \ge z_{j_3} \quad ((j_1,j_2,j_3) \in H), \label{con3:MIQO3}\\
& \quad z_j \in \{0,1\} \quad (j \in G). \label{con4:MIQO3}
\end{align}

\section{Computational experiments}\label{sec:4}
In this section, we assess the effectiveness of our MIO models for hierarchical variable selection and examine the store-choice factors that are inferred from the computational results.

\subsection{Datasets}
We used scanner panel data that were provided by the Japanese marketing research company MACROMILL (\url{http://www.macromill.com/global/index.html}). 
These data were collected from home scans by roughly 4,000 consumers during 2012--2013. We focused on choosing among three types of stores in Tokyo: convenience stores, drugstores, and grocery supermarkets. Each store type comprised the leading five chains as determined by sales volume, as given in Table~\ref{tab:store}. 
\begin{table}
\caption{Store types and chains}
\label{tab:store}
\begin{tabular}{ll}
\hline\noalign{\smallskip}
Store type & Store chains \\
\noalign{\smallskip}\hline\noalign{\smallskip}
Convenience store   & 7-Eleven, Lawson, FamilyMart, Ministop, Three-F \\
Drugstore           & Matsumotokiyoshi, Sundrug, Tsuruha, Cosmos, Sugi \\
Grocery supermarket & Aeon, Seven\&I, Uny, Daiei, Izumi \\
\noalign{\smallskip}\hline
\end{tabular}
\end{table}

Each sample corresponded to one consumer visit and was designated as either a positive or negative example according to which store was visited. The three datasets that we analyzed are listed in Table~\ref{tab:dataset}, where $n$ is the number of samples.
\begin{table}
\caption{Datasets on store choice}
\label{tab:dataset}
\begin{tabular}{lllr}
\hline\noalign{\smallskip}
Abbreviation & Positive example ($y_i > 0$) & Negative example ($y_i < 0$) & $n$ \\
\noalign{\smallskip}\hline\noalign{\smallskip}
CvsD & convenience stores & drugstores           & 225,630 \\
CvsS & convenience stores & grocery supermarkets & 252,513 \\
DvsS & drugstores         & grocery supermarkets & 139,188 \\
\noalign{\smallskip}\hline
\end{tabular}
\end{table}

As shown in Table~\ref{tab:cnsmrs}, 37 dummy consumer-demographic variables were created from the results of a questionnaire survey of consumers. We prepared dummy variables corresponding to 5 large categories, 29 medium categories, and 214 small categories of products. These were based on the Japan Item Code File Service (JICFS), which is a standard means of classifying products in Japan. For reference, the large and medium product categories are listed in Table~\ref{tab:ctgrs}. Note that any redundant variable (i.e., zero in all samples) was eliminated in each dataset, which is why the number of candidate explanatory variables differs among datasets.
\begin{table}
\caption{Consumer demographics}
\label{tab:cnsmrs}
\begin{tabular}{ll}
\hline\noalign{\smallskip}
Consumer demographic & Explanatory variables \\
\noalign{\smallskip}\hline\noalign{\smallskip}
Gender/family & female, married, with~children \\ \noalign{\smallskip}
Age           & 20s, 30s, 40s, 50s, over~60s \\ \noalign{\smallskip}
Income        & class~2~or~lower, classes~3--4, classes~5--6, classes~7--9, \\            & class~10~or~higher \\\noalign{\smallskip}
Cross terms   & $\rm\{male, female\}\times\{married, with~children\}$ \\
              & $\rm\{male, female\}\times age$ \\
              & $\rm\{male, female\}\times income~class$ \\
\noalign{\smallskip}\hline
\end{tabular}
\end{table}
\begin{table}
\caption{Large and medium categories of products}
\label{tab:ctgrs}
\begin{tabular}{ll}
\hline\noalign{\smallskip}
Large category & Medium categories \\
\noalign{\smallskip}\hline\noalign{\smallskip}
Food                       & Processed foods, fresh foods, confectioneries, beverages, \\
                           & other foods \\ \noalign{\smallskip}
Commodities                & Everyday sundries, healthcare, cosmetics, housewares, \\                         & DIY supplies, pet care/food, other commodities \\ \noalign{\smallskip}
Cultural supplies          & Stationery/office supplies, toys, books, musical instruments, \\
                           & information equipment, other cultural supplies \\ \noalign{\smallskip}
Durables                   & Furniture, car supplies, watches/clocks/glasses, \\
                           & optics/photo, home electronics, other durables \\ \noalign{\smallskip}
Clothes/accessories/sports & Clothes, bedding, accessories, footwear, sports equipment \\
\noalign{\smallskip}\hline
\end{tabular}
\end{table}

In the following sections, we use two sets of candidate explanatory variables: 
\begin{description}
\item[\bf C-LM set:] set of explanatory variables of consumer demographics and \{large, medium\} product categories; 
\item[\bf C-LMS set:] set of explanatory variables of consumer demographics and \{large, medium, small\} product categories. 
\end{description}
Note that the small product categories were excluded from the C-LM set. 

\subsection{Evaluation of predictive performance}
We evaluated the predictive performance of each of the following methods for variable selection by means of five-fold cross-validation: 
\begin{description}
\item[\bf SW] stepwise method~(\citealt{Ef60})
\item[\bf L1] $L_1$-regularized regression~(\citealt{Ti96})
\item[\bf BM] basic MIO model~\eqref{obj:MIQO1}--\eqref{con3:MIQO1}~(\citealt{BeKi16b,KoYa09})
\item[\bf SHM] strong hierarchical MIO model~\eqref{obj:MIQO2}--\eqref{con4:MIQO2}
\item[\bf WHM] weak hierarchical MIO model~\eqref{obj:MIQO3}--\eqref{con4:MIQO3}
\end{description}

All computations were performed on a Macintosh computer with an Intel Xeon X5650 CPU ($2\times2.66$~GHz) and 64~GB of memory. The stepwise method was started with no explanatory variables, whereupon the variable that led to the largest decrease in Akaike's information criterion was added or eliminated iteratively until $s$ variables had been selected. This was performed using the {\tt step} function in R 3.2.0~(\url{http://www.R-project.org}). In the case of $L_1$-regularized regression, the regression coefficients were estimated using the {\tt glmnet} package in R 3.2.0 for each value of the regularization parameter chosen from $\{0, 0.0001, 0.0002, \ldots, 0.9999, 1\}$. We then chose a set of $s$ variables that had non-zero coefficients, and we estimated those coefficients again using the ordinary least-squares method. The MIO problems were solved using Gurobi Optimizer 6.5~(\url{http://www.gurobi.com}). Here, the MIO computation time was limited to 1,000\,s; that is, if a computation had not finish by itself within 1,000\,s, the best feasible solution obtained thus far was used as the result. 

Tables~\ref{tab:resultsCLM} and \ref{tab:resultsCLMS} give the results of five-fold cross-validation for the C-LM and C-LMS sets, respectively. Here, there are $p$ candidate explanatory variables in each dataset, and $s$ is the upper bound on the number of selected explanatory variables, where $s = 10,20$ for the C-LM set and $s = 10,20,50$ for the C-LMS set. The columns labeled ``$R^2$'' and ``RMSE'' are the average values of the coefficient of determination and the root-mean-square error measured through the five-fold cross-validation, respectively. These quantify the predictive performance, and the best values among the five methods are indicated in bold. The column labeled ``Time~(s)'' is the average computation time in seconds required for variable selection. 

We begin by focusing on the results for the C-LM set (Table~\ref{tab:resultsCLM}), from which the small product categories were excluded. In this case, SHM and WHM are the same model, so these results are shown as ``S/WHM'' in the table. We can see that although SW, BM, and S/WHM provided relatively good predictive performance, the average $R^2$ and RMSE values of S/WHM were the best among all the methods. The worst predictive performance was always delivered by L1 . In contrast to the other methods, S/WHM exploited the hierarchical relationships of product categories in the course of variable selection. As a result, its predictive performance was better than those of the others. 

\begin{table}
\caption{Results of five-fold cross-validation for C-LM set}
\label{tab:resultsCLM}
\begin{tabular}{lrrlrrr}
\hline\noalign{\smallskip}
Dataset & $p$ & $s$ & Method & $R^2$ & RMSE & Time~(s) \\
\noalign{\smallskip}\hline\noalign{\smallskip}
CvsD & 69 & 10 &
   SW    & \textbf{0.3367} & \textbf{2.7090} &   106.4 \\
&&&L1    & 0.3360          & 2.7104          &    10.0 \\
&&&BM    & \textbf{0.3367} & \textbf{2.7090} &    43.3 \\
&&&S/WHM & 0.3366          & 2.7091          &    14.9 \\ \noalign{\smallskip}
&& 20 & 
   SW    & \textbf{0.3384} & \textbf{2.7055} &   311.1 \\
&&&L1    & 0.3375          & 2.7073          &     6.8 \\
&&&BM    & 0.3381          & 2.7060          & 1,000.0 \\
&&&S/WHM & 0.3382          & 2.7059          & 1,000.0 \\ \noalign{\smallskip}
CvsS & 71 & 10 & 
   SW    & 0.2105          & 5.2533          &   121.9 \\
&&&L1    & 0.2090          & 5.2583          &    24.7 \\
&&&BM    & 0.2105          & 5.2533          &    43.7 \\
&&&S/WHM & \textbf{0.2106} & \textbf{5.2530} &    14.4 \\ \noalign{\smallskip}
&& 20 & 
   SW    & 0.2140          & 5.2416          &   360.5 \\
&&&L1    & 0.2138          & 5.2426          &    14.5 \\
&&&BM    & 0.2138          & 5.2423          & 1,000.0 \\
&&&S/WHM & \textbf{0.2143} & \textbf{5.2406} & 1,000.0 \\ \noalign{\smallskip}
DvsS & 71 & 10 & 
   SW    & \textbf{0.2098} & \textbf{6.8622} &    66.4 \\
&&&L1    & 0.2089          & 6.8663          &    28.6 \\
&&&BM    & 0.2097          & 6.8626          &    18.0 \\
&&&S/WHM & 0.2097          & 6.8628          &     3.9 \\ \noalign{\smallskip}
&& 20 & 
   SW    & 0.2113          & 6.8555          &   195.8 \\
&&&L1    & 0.2110          & 6.8570          &    13.1 \\
&&&BM    & 0.2114          & 6.8552          & 1,000.0 \\
&&&S/WHM & \textbf{0.2116} & \textbf{6.8545} & 1,000.0 \\ \noalign{\smallskip}
average&&&
   SW    & 0.2535          & 4.9379          &   193.7 \\
&&&L1    & 0.2527          & 4.9403          &    16.3 \\
&&&BM    & 0.2534          & 4.9381          &   517.5 \\
&&&S/WHM & \textbf{0.2535} & \textbf{4.9377} &   505.5 \\ 
\noalign{\smallskip}\hline
\end{tabular}
\end{table}

We next discuss the results for the C-LMS set (Table~\ref{tab:resultsCLMS}), in which the small product categories were included. As in Table~\ref{tab:resultsCLM}, good predictive performance was achieved by SW, SHM, and WHM, whereas the average $R^2$ and RMSE values were slightly better for WHM than for the others. In contrast, the predictive performance of BM was greatly decreased. The main reason for this was that the C-LMS set involved many candidate explanatory variables, and thus BM failed to provide quality solutions because of the limited computation time. Nevertheless, SHM and WHM maintained high predictive performance for the C-LMS set because the number of feasible solutions was reduced by the hierarchical constraints. 

The best predictive performance for the CvsD dataset with $s = 50$ was attained by SW (Table~\ref{tab:resultsCLMS}), but it is noteworthy that SW took more than 6,000\,s to compute. If we quit the stepwise method in the middle of computation, the number of selected variables did not reach $s$. Because such a shortage of selected variables could lead to an unfair comparison, we did not set a time limit on the SW computation. However, we should note that the computation time of SW was 1,047.0\,s for the CvsD dataset with $s = 20$ (Table~\ref{tab:resultsCLMS}). In other words, SW would select fewer than 20 variables if the computation time is limited to 1,000\,s. In addition, the predictive performance was much lower for SW with $s = 20$ than for SHM and WHM with $s = 50$. In fact, for the CvsD dataset, the $R^2$ value of SW with $s = 20$ was 0.3642, and those of SHM and WHM with $s = 50$ were 0.3693 and 0.3697, respectively. Taking all aspects into consideration, it is clear that SW was inferior to SHM and WHM in relation to the selection of many explanatory variables within a limited time. In contrast, although L1 delivered relatively low predictive performance, its computation was extremely rapid. 

\begin{table}
\caption{Results of five-fold cross-validation for C-LMS set}
\label{tab:resultsCLMS}
\begin{tabular}{lrrlrrr}
\hline\noalign{\smallskip}
Dataset & $p$ & $s$ & Method & $R^2$ & RMSE & Time~(s) \\
\noalign{\smallskip}\hline\noalign{\smallskip}
CvsD & 234 & 10 & 
   SW  & \textbf{0.3531} & \textbf{2.6752} &   348.2 \\
&&&L1  & 0.3467          & 2.6886          &    10.1 \\
&&&BM  & 0.3346          & 2.7132          & 1,000.0 \\
&&&SHM & 0.3509          & 2.6798          & 1,000.0 \\
&&&WHM & 0.3530          & 2.6755          & 1,000.0 \\
\noalign{\smallskip}
&& 20 & 
   SW  & 0.3642          & 2.6522          & 1,047.0 \\
&&&L1  & 0.3612          & 2.6586          &     2.7 \\
&&&BM  & 0.3412          & 2.6997          & 1,000.0 \\
&&&SHM & 0.3618          & 2.6573          & 1,000.0 \\
&&&WHM & \textbf{0.3644} & \textbf{2.6518} & 1,000.0 \\
\noalign{\smallskip}
&& 50 & 
   SW  & \textbf{0.3701} & \textbf{2.6400} & 6,018.8 \\
&&&L1  & 0.3691          & 2.6422          &    17.5 \\
&&&BM  & 0.3552          & 2.6708          & 1,000.0 \\
&&&SHM & 0.3693          & 2.6417          & 1,000.0 \\
&&&WHM & 0.3697          & 2.6408          & 1,000.0 \\
\noalign{\smallskip}
CvsS & 277 & 10 & 
   SW  & \textbf{0.2340} & \textbf{5.1751} &   551.9 \\
&&&L1  & 0.2265          & 5.2004          &    22.6 \\
&&&BM  & 0.1671          & 5.3965          & 1,000.0 \\
&&&SHM & 0.2330          & 5.1786          & 1,000.0 \\
&&&WHM & \textbf{0.2340} & \textbf{5.1751} & 1,000.0 \\
\noalign{\smallskip}
&& 20 & 
   SW  & 0.2440          & 5.1413          & 1,439.6 \\
&&&L1  & 0.2369          & 5.1655          &    15.0 \\
&&&BM  & 0.2004          & 5.2865          & 1,000.0 \\
&&&SHM & \textbf{0.2486} & \textbf{5.1256} & 1,000.0 \\
&&&WHM & 0.2441 & 5.1412                   & 1,000.0 \\
\noalign{\smallskip}
&& 50 & 
   SW  & 0.2511          & 5.1172          & 8,331.7 \\
&&&L1  & 0.2495          & 5.1228          &    29.9 \\
&&&BM  & 0.2314          & 5.1838          & 1,000.0 \\
&&&SHM & \textbf{0.2511} & \textbf{5.1170} & 1,000.0 \\
&&&WHM & 0.2511          & 5.1172          & 1,000.0 \\
\noalign{\smallskip}
DvsS & 281 & 10 & 
   SW  & \textbf{0.2199} & \textbf{6.8176} &   256.4 \\
&&&L1  & 0.2179          & 6.8263          &    23.9 \\
&&&BM  & 0.1916          & 6.9399          & 1,000.0 \\
&&&SHM & 0.2176          & 6.8278          & 1,000.0 \\
&&&WHM & \textbf{0.2199} & \textbf{6.8176} & 1,000.0 \\
\noalign{\smallskip}
&& 20 & 
   SW  & 0.2256          & 6.7927          &   786.7 \\
&&&L1  & 0.2243          & 6.7984          &    10.1 \\
&&&BM  & 0.2088          & 6.8658          & 1,000.0 \\
&&&SHM & 0.2255          & 6.7930          & 1,000.0 \\
&&&WHM & \textbf{0.2264} & \textbf{6.7889} & 1,000.0 \\
\noalign{\smallskip}
&& 50 & 
   SW  & 0.2286          & 6.7795          & 4,551.5 \\
&&&L1  & \textbf{0.2289} & \textbf{6.7782} &    21.9 \\
&&&BM  & 0.2228          & 6.8045          & 1,000.0 \\
&&&SHM & 0.2289          & 6.7782          & 1,000.0 \\
&&&WHM & 0.2286          & 6.7792          & 1,000.0 \\
\noalign{\smallskip}
average&&&
   SW  & 0.2767          & 4.8656          & 2,592.4 \\
&&&L1  & 0.2734          & 4.8757          &    17.1 \\
&&&BM  & 0.2503          & 4.9512          & 1,000.0 \\
&&&SHM & 0.2763          & 4.8666          & 1,000.0 \\
&&&WHM & \textbf{0.2768} & \textbf{4.8653} & 1,000.0 \\ 
\noalign{\smallskip}\hline
\end{tabular}
\end{table}

\subsection{Analysis of store-choice factors}
Tables~\ref{tab:varCvsD}--\ref{tab:varDvsS} give the explanatory variables selected by WHM with $s = 20$ for the C-LMS set. Here, all the samples in each dataset were used for variable selection, and the time limit for an MIO computation was extended to 10,000\,s. In the tables, (L), (M), and (S) stand for large, medium, and small product categories, respectively. 

\subsubsection{Convenience store versus drugstore}

Table~\ref{tab:varCvsD} gives the results for the CvsD dataset, for which the variables with positive coefficients were the choice factors of convenience stores, and those with negative coefficients were the choice factors of drugstores. The absolute value of a coefficient indicates the number of products purchased together. For instance, in Table~\ref{tab:varCvsD} we see that ``other commodities~(M)'' was a choice factor of convenience stores, and that the purchase of ``other commodities~(M)'' led to the purchase of roughly five products from such a store. 

In the case of WHM, if a certain category is selected, at least one of its superordinate categories has to be selected because of the weak hierarchical constraints. For instance, Table~\ref{tab:varCvsD} shows that ``other foods~(M)'' was selected along with ``food~(L),'' and that ``sanitary papers~(S),'' ``everyday sundries~(M),'' and ``commodities~(L)'' were selected according to the hierarchical relationship. Meanwhile, these variables can have coefficients with opposite signs. In fact, the coefficients of ``sanitary papers~(S)'' and ``commodities~(L)'' were $1.09$ and $-4.67$, respectively. 
In what follows, we analyze these results in relation to product price, store accessibility, and selection of products. 

\begin{table}
\caption{Explanatory variables selected by WHM in CvsD dataset 
(positive example: convenience store; negative example: drugstore)}
\label{tab:varCvsD}
\begin{tabular}{rrl}
\hline\noalign{\smallskip}
Explanatory variable & \multicolumn{2}{c}{Coefficient} \\
\noalign{\smallskip}\hline\noalign{\smallskip}
other commodities~(M)          & $5.38$  & *** \\
nutritional fortification~(S)  & $3.82$  & *** \\
other housewares~(S)           & $2.82$  & *** \\
intercept term                 & $2.71$  & *** \\
other foods~(S)                & $2.30$  & *** \\
hygiene/medical care~(S)       & $1.40$  & *** \\
processed meats~(S)            & $1.38$  & *** \\
delicacies~(S)                 & $1.19$  & *** \\
frozen foods~(S)               & $1.19$  & *** \\
sanitary papers~(S)            & $1.09$  & *** \\
alcohol~(S)                    & $0.91$  & *** \\
delicatessen~(S)               & $0.86$  & *** \\
ice cream~(S)                  & $0.74$  & *** \\
female$\times$with children    & $-0.45$ & *** \\
food~(L)                       & $-1.17$ & *** \\
tofu/konjac~(S)                & $-1.23$ & *** \\
everyday sundries~(M)          & $-1.27$ & *** \\
durables~(L)                   & $-1.57$ & *** \\
other foods~(M)                & $-1.67$ & *** \\
clothes/accessories/sports~(L) & $-2.44$ & *** \\
commodities~(L)                & $-4.67$ & *** \\
\noalign{\smallskip}\hline
\end{tabular}\\
***: $p$-value $<$ 0.001
\end{table}

In Table~\ref{tab:varCvsD}, ``commodities~(L)'' was a choice factor of drugstores, but ``other commodities~(M)'' was a choice factor of convenience stores. The choice factor ``commodities~(L)'' contained featured products of drugstores, such as tissues, toilet paper, and detergent. Hence, consumers tended to purchase these products at drugstores because of their lower prices. In contrast, ``other commodities~(M)'' was composed largely of Amazon/iTunes gift cards and garbage disposal permits from convenience stores. These products were purchased in convenience stores probably because of their accessibility. Additionally, the coefficient of ``other commodities~(M)'' was very large (i.e., 5.38); that is, when those products were purchased, other products were likely to be purchased at the same time. In contrast, food categories such as ``frozen foods~(S),'' ``alcohol~(S),'' ``delicatessen~(S),'' and ``ice cream~(S)'' had relatively small positive coefficients. This means that consumers would purchase only those products in a single visit, and hence would choose convenience stores because of their accessibility. 

The choice factor ``everyday sundries~(M)'' was composed primarily of detergents, tissues, and toilet paper purchased from drugstores, probably because of the lower prices. The choice factor ``clothes/accessories/sports~(L)'' consisted mainly of women's beauty products that are specific to drugstores, such as stockings, tights, and open-toe socks. Moreover, because ``female$\times$with children'' and ``commodities~(L)'' were choice factors of drugstores, we reason that it is on economic grounds that women shop at drugstores for commodities for their families.

\subsubsection{Convenience store versus grocery supermarket}
Table~\ref{tab:varCvsS} gives the results for the CvsS dataset, for which the variables with positive coefficients were the choice factors of convenience stores, and those with negative coefficients were the choice factors of grocery supermarkets. As in Table~\ref{tab:varCvsD}, the coefficient of ``other commodities~(M)'' was a very large positive value (i.e., 6.63). Accordingly, gift cards and garbage disposal permits were still major choice factors of convenience stores. Other choice factors of convenience stores included ``bread/cereal~(S),'' ``soup~(S),'' ``delicatessen~(S),'' ``noodles~(S),'' and ``frozen foods~(S).'' These easy-to-prepare products were purchased at convenience stores probably because of accessibility. In contrast, the choice factors of grocery supermarkets included ``male$\times$over~60s,'' ``female$\times$married,'' and ``female$\times$with~children'' as consumer demographics, and ``dessert/yogurt~(S),'' ``milk~beverage~(S),'' and ``fresh~foods~(M)'' as product categories. In other words, the typical customers of grocery supermarkets were elderly men and married women who purchased healthy foods there. The large categories ``food~(L),'' ``durables~(L),'' and ``commodities~(L)'' were also choice factors of grocery supermarkets; these are typical products of general merchandise stores and shopping malls. 

\begin{table}
\caption{Explanatory variables selected by WHM in CvsS dataset 
(positive example: convenience store; negative example: grocery supermarket)}
\label{tab:varCvsS}
\begin{tabular}{rrl}
\hline\noalign{\smallskip}
Explanatory variable & \multicolumn{2}{c}{Coefficient} \\
\noalign{\smallskip}\hline\noalign{\smallskip}
other commodities~(M)         &  $6.63$ & *** \\
nutritional fortification~(S) &  $6.50$ & *** \\
bread/cereal~(S)              &  $4.59$ & *** \\
soup~(S)                      &  $3.70$ & *** \\
delicatessen~(S)              &  $3.28$ & *** \\
intercept term                &  $3.06$ & *** \\
hygiene/medical care~(S)      &  $3.05$ & *** \\
noodles~(S)                   &  $2.53$ & *** \\
frozen foods~(S)              &  $1.73$ & *** \\
other foods~(S)               &  $1.02$ & *** \\
male$\times$over 60s          & $-0.59$ & *** \\
female$\times$married         & $-0.78$ & *** \\
dessert/yogurt~(S)            & $-0.92$ & *** \\
milk beverage~(S)             & $-1.22$ & *** \\
female$\times$with children   & $-1.54$ & *** \\
food~(L)                      & $-1.71$ & *** \\
fresh foods~(M)               & $-2.05$ & *** \\
coffee/tea~(S)                & $-2.20$ & *** \\
durables~(L)                  & $-2.87$ & *** \\
commodities~(L)               & $-5.61$ & *** \\
processed foods~(M)           & $-5.78$ & *** \\
\noalign{\smallskip}\hline
\end{tabular}\\
***: $p$-value $<$ 0.001
\end{table}

\subsubsection{Drugstore versus grocery supermarket}
Table~\ref{tab:varDvsS} gives the results for the DvsS dataset, for which the variables with positive coefficients were the choice factors of drugstores, and those with negative coefficients were the choice factors of grocery supermarkets. We can see that drugstores were chosen for typical product categories such as ``health foods~(S),'' ``infant foods~(S),'' ``healthcare~(M),'' and ``cosmetics~(M).'' Additionally, ``bread/cereal~(S),'' ``soup~(S),'' and ``noodles~(S)'' were common choice factors of convenience stores in Table~\ref{tab:varCvsS} and drugstores in Table~\ref{tab:varDvsS}. In other words, convenience stores and drugstores were chosen for similar products when compared to grocery supermarkets. In addition, some food products such as ``health foods~(S),'' ``infant foods~(S),'' ``bread/cereal~(S),'' and ``grain~(S)'' had large positive coefficients. Hence, these products could stimulate impulse purchases at drugstores and so increase average customer sales. In contrast, the large categories ``commodities~(L)'' and ``food~(L)'' were choice factors of grocery supermarkets, which were chosen especially for ``processed foods~(M)'' by ``female$\times$with~children.'' To summarize the results of Tables~\ref{tab:varCvsS} and \ref{tab:varDvsS}, grocery supermarkets should provide a good selection of health food products for women with families. 

\begin{table}
\caption{Explanatory variables selected by WHM in DvsS dataset 
(positive example: drugstore; negative example: grocery supermarket)}
\label{tab:varDvsS}
\begin{tabular}{rrl}
\hline\noalign{\smallskip}
Explanatory variable & \multicolumn{2}{c}{Coefficient} \\
\noalign{\smallskip}\hline\noalign{\smallskip}
health foods~(S)            & $5.42$  & *** \\
infant foods~(S)            & $4.00$  & *** \\
bread/cereal~(S)            & $3.48$  & *** \\
grain~(S)                   & $3.39$  & *** \\
healthcare~(M)              & $3.15$  & *** \\
everyday sundries~(M)       & $2.60$  & *** \\
cosmetics~(M)               & $2.50$  & *** \\
soup~(S)                    & $2.14$  & *** \\
sweets~(S)                  & $2.03$  & *** \\
soft drink~(S)              & $1.62$  & *** \\
noodles~(S)                 & $1.59$  & *** \\
tofu/konjac~(S)             & $1.35$  & *** \\
seasoning~(S)               & $1.32$  & *** \\
female$\times$under~class~2 & $0.67$  & *** \\
intercept term              & $0.08$  & \\
commodities~(L)             & $-0.42$ & *** \\
female$\times$with~children & $-0.61$ & *** \\
other~housewares~(S)        & $-2.64$ & *** \\
cultural supplies~(L)       & $-2.66$ & *** \\
food~(L)                    & $-3.07$ & *** \\
processed foods~(M)         & $-4.88$ & *** \\
\noalign{\smallskip}\hline
\end{tabular}\\
***: $p$-value $<$ 0.001
\end{table}

\subsubsection{Intercept term}
We conclude this section by discussing the intercept term, which can be interpreted as quantifying a store's ability to attract consumers once the effects of all other consumer/product attributes have been eliminated. We can see from Tables~\ref{tab:varCvsD} and \ref{tab:varCvsS} that convenience stores had the highest potential to attract consumers. This suggests that impulse purchases would be made frequently at convenience stores.

\section{Conclusions}\label{sec:5}
We proposed an MIO-based method of hierarchical variable selection and analyzed consumer store-choice factors based on purchase quantity. Our method improved the predictive performance based on hierarchically structured product categories. It also offered better computational efficiency because the hierarchical constraints reduced the number of feasible solutions appreciably. We verified the effectiveness of our method by comparing its computational performance with those of the stepwise method and $L_1$-regularized regression through a five-fold cross-validation. 

We used our variable selection method to examine consumer store choice among convenience stores, drugstores, and grocery supermarkets. We found from the analysis that convenience stores were chosen because of their accessibility, whereas drugstores were chosen in order to purchase specific products at low prices. We also found that grocery supermarkets were chosen especially for health food products and were favored by women with families.

A future direction of study would be to append constraints for eliminating multicollinearity~(\citealt{BeKi16a,TaKo16,TaKo17}) to our MIO models. These constraints could further improve the accuracy of coefficient estimates in the store-choice model. Although our computational experiments were focused on three store types, various analyses of store-choice factors could be performed using our MIO models. For instance, consumer choice between stores of the same type could be investigated, and seasonal/regional store-choice factors could be identified by applying our store-choice model to the data of each season/region. To gain customers in today's fiercely competitive market, each store/company needs to accelerate its product development, have a good selection of products, and examine how it sets product prices. Accordingly, the results of our store-choice analysis would be useful in understanding the strengths and weaknesses of each store, and for devising its unique marketing strategy.

\if0
\section{Introduction}
\label{intro}
Your text comes here. Separate text sections with
\section{Section title}
\label{sec:1}
Text with citations \cite{RefB} and \cite{RefJ}.
\subsection{Subsection title}
\label{sec:2}
as required. Don't forget to give each section
and subsection a unique label (see Sect.~\ref{sec:1}).
\paragraph{Paragraph headings} Use paragraph headings as needed.
\begin{equation}
a^2+b^2=c^2
\end{equation}

% For one-column wide figures use
\begin{figure}
% Use the relevant command to insert your figure file.
% For example, with the graphicx package use
  \includegraphics{example.eps}
% figure caption is below the figure
\caption{Please write your figure caption here}
\label{fig:1}       % Give a unique label
\end{figure}
%
% For two-column wide figures use
\begin{figure*}
% Use the relevant command to insert your figure file.
% For example, with the graphicx package use
  \includegraphics[width=0.75\textwidth]{example.eps}
% figure caption is below the figure
\caption{Please write your figure caption here}
\label{fig:2}       % Give a unique label
\end{figure*}
%
% For tables use
\begin{table}
% table caption is above the table
\caption{Please write your table caption here}
\label{tab:1}       % Give a unique label
% For LaTeX tables use
\begin{tabular}{lll}
\hline\noalign{\smallskip}
first & second & third  \\
\noalign{\smallskip}\hline\noalign{\smallskip}
number & number & number \\
number & number & number \\
\noalign{\smallskip}\hline
\end{tabular}
\end{table}
\fi

\begin{acknowledgements}
%If you'd like to thank anyone, place your comments here
%and remove the percent signs.
This research was partially supported by a Grant-in-Aid of Joint Research from the Institute of Information Science, Senshu University. 
\end{acknowledgements}

% BibTeX users please use one of
%\bibliographystyle{spbasic}      % basic style, author-year citations
%\bibliographystyle{spmpsci}      % mathematics and physical sciences
%\bibliographystyle{spphys}       % APS-like style for physics
%\bibliography{}   % name your BibTeX data base

% Non-BibTeX users please use

\end{document}